\pdfoutput=1

\documentclass[openacc]{rstransa}
\usepackage{graphicx, amsmath, amssymb, nicefrac}
\usepackage[normalem]{ulem}
\usepackage{enumitem}

\usepackage{bm}  

\newcommand{\CaIR}{Ca~\textsc{ii}~8542{\,}{\AA}}

\usepackage{tcolorbox}
\definecolor{abstractcolor}{cmyk}{.04,.04,.12,.08}

\begin{document}

\title{Accurately constraining velocity information from spectral imaging observations using machine learning techniques}

\author{
Conor D. MacBride$^{1}$, David B. Jess$^{1,2}$, Samuel D. T. Grant$^{1}$, Elena Khomenko$^{3,4}$, Peter H. Keys$^{1}$ and Marco Stangalini$^{5}$}

\address{$^{1}$Astrophysics Research Centre, School of Mathematics and Physics, Queen's University Belfast, Belfast, BT7 1NN, UK \\
$^{2}$Department of Physics and Astronomy, California State University Northridge, Northridge, CA 91330, U.S.A.\\
$^{3}$Instituto de Astrof{\'{i}}sica de Canarias, 38205 La Laguna, Tenerife, Spain\\
$^{4}$Departamento de Astrof{\'{i}}sica, Universidad de La Laguna, 38205 La Laguna, Tenerife, Spain\\
$^{5}$Italian Space Agency (ASI), Via del Politecnico snc, 00133 Roma, Italy\\}

\subject{artificial intelligence, astrophysics, computational physics, observational astronomy, pattern recognition, solar system, spectroscopy}


\corres{Conor D. MacBride\\
\email{cmacbride01@qub.ac.uk}}

\begin{abstract}
 Determining accurate plasma Doppler (line-of-sight) velocities from spectroscopic measurements is a challenging endeavour, especially when weak chromo-spheric absorption lines are often rapidly evolving and, hence, contain multiple spectral components in their constituent line profiles. Here, we present a novel method that employs machine learning techniques to identify the underlying components present within observed spectral lines, before subsequently constraining the constituent profiles through single or multiple Voigt fits. Our method allows active and quiescent components present in spectra to be identified and isolated for subsequent study. Lastly, we employ a {\CaIR} spectral imaging dataset as a proof-of-concept study to benchmark the suitability of our code for extracting two-component atmospheric profiles that are commonly present in sunspot chromospheres. Minimisation tests are employed to 
\end{abstract}

\maketitle

\begin{tcolorbox}[sharp corners, width=\textwidth,colback=abstractcolor,colframe=abstractcolor,boxsep=5pt,left=0pt,right=0pt,top=0pt,bottom=0pt]
	validate the reliability of the results, achieving median reduced $\chi^2$ values equal to 1.03 between the observed and synthesised umbral line profiles.
\end{tcolorbox}

\section{Introduction}
\label{sec:introduction}

Tomographic analysis of the complex solar atmosphere and its dynamics requires the combined use of different spectral lines (e.g. Ca~{\sc{ii}}, Mg~{\sc{ii}}, L$\alpha$, and H$\alpha$).
Indeed, due to radiative transfer effects, these lines are formed across different heights of the solar atmosphere, with spectral information extracted at a particular percentage line depth corresponding to differing optical depths, and hence particular atmospheric heights \cite{Vernazza1981}.
Hence, it is possible to extract solar plasma properties as a function of optical depth (and atmospheric height if radiative transfer processes are well constrained) by taking measurements at specific percentage line depths \cite{GonzalezManrique2020}.

The challenge of the solar chromosphere is that non-local thermodynamic equilibrium (non-LTE) plays an important role in the propagation of photons through the relatively dense and tenuous atmosphere \cite{SocasNavarro2015}. Furthermore, due to spatial resolution limitations of even the largest-aperture ground-based solar telescopes, and the intrinsic fine-scale structuring of the solar atmosphere (scales smaller than $100-150$~km), often the observed spectrum within the resolution element is the superposition of several components owing to different plasma states (e.g., both quiescent and magnetised atmospheric components).
The effects of  a multi-component atmosphere are generally taken into account in state-of-the-art spectropolarimetric inversion codes (e.g. NICOLE; \cite{SocasNavarro2015}).
However, these techniques are extremely demanding computationally. 
In addition to the effects of spatial resolution, multi-components can arise from the fact that chromospheric spectral lines sample a wide range of heights; from the mid-photosphere (wings of the line), through to the upper chromosphere (core of the line), thus mixing very different physical regimes \cite{Khomenko2015}.

Additional spectroscopic components that are superimposed on top of the quiescent background spectral profile can arise from a wealth of solar phenomena, including dynamic events such as magnetic reconnection, propagating waveforms, and shock development, all of which have the ability to form across a range of atmospheric heights.
This adds a secondary component to an otherwise quiescent atmosphere \cite{Felipe2014}, which is often highly Doppler-shifted relative to the approximately stationary line core, creating a significantly broadened resultant spectral profile.
Being able to segregate the multi-component atmospheric contributions from a spectral line would allow the simultaneous examination of both `dynamic' and `quiescent' regimes of the atmosphere much more readily. 

One of the most dynamic and widely recognised processes that creates multi-component chromospheric spectral lines is the propagation of magnetoacoustic waves, and their subsequent development into shock fronts, in the umbrae of sunspots \cite{Beckers1969,RouppeVanDerVoort2003,Centeno2006,delaCruzRodriguez2013b,Henriques2017,Houston2018}.
Such umbral oscillations propagate energy upwards from the photosphere into the chromosphere, where the steep density drop results in the rapid increase of the velocity amplitude in an attempt to conserve energy flux, which is readily captured by the temperature-sensitive {\CaIR} spectral line. 
Once the velocity amplitude exceeds the local sound speed, the waves develop into shock fronts -- a process that can occur as low as $\sim$250~km \cite{Grant2018} through to the uppermost region of the chromosphere \cite{Houston2018}, resulting in the production of optically thin {\CaIR} emission \cite{Lites1984}.
It is this optically thin emission that superimposes on top of the quiescent {\CaIR} spectral profile, appearing in spectroscopic observations as an almost instantaneous blue-shifted (i.e., upwardly propagating) emission that slowly decays from its maximum intensity, providing the characteristic `saw-tooth' spectral shape that is synonymous with umbral shock formation.
Such spectral profile evolution is coupled to the underlying $p$-mode wave spectrum, providing a periodicity of approximately 3~minutes to the umbral flashes \cite{Louis2014, SocasNavarro2000}. 
Attempting to fit a multi-component umbral flash spectral profile using centre-of-gravity methods \cite{Rees1979, Uitenbroek2003} or a single Gaussian, Lorentzian, or Voigt profile is prone to error, since it may vastly under- or over-estimate the plasma characteristics of the developing shock.
Hence, to fully understand the physics behind rapid atmospheric evolution (including those related to umbral flashes), both components must be considered separately; not simply modelled as a single-component atmosphere that attempts to bridge quiescent and active states \cite{SocasNavarro2000, Grant2018}.

Doppler (line-of-sight) velocities are typically acquired from spectroscopic observations using a number of techniques.
They can be measured across a range of optical depths by calculating, at specific percentage line depths, the spectral line bisector shift from the central line core wavelength \cite{GonzalezManrique2020,Kulander1966}.
If full spectropolarimetric measurements are available, other atmospheric parameters (such as temperature, line-of-sight magnetic flux, and magnetic field inclination angles) can be estimated by inversion codes \cite{Keys2020}.
Such high-precision measurements of Stokes~$I/Q/U/V$ are difficult to obtain for deep chromospheric absorption lines, including {\CaIR}, due to the relatively small rate at which photons reach the detector\cite{Jess2015, Stangalini2018}.  
To combat the inherently weak signal, exposure times could be increased accordingly, but this is not always feasible when tracking the evolution of dynamic and rapidly evolving features in the lower solar atmosphere \cite{Felipe2018}.
If only Stokes~$I$ spectral observations are available, Voigt profiles \cite{Zaghloul2007, Jess2019} can be fitted to the spectra.
As we will discuss in Section~\ref{sec:methods}, the Voigt profile includes enhanced spectral wing broadening, which provides a more suitable realisation of the radiative transfer effects omnipresent throughout the lower solar atmosphere, including Doppler and pressure broadening mechanisms \cite{Shine1972,Cauzzi2008,delaCruzRodriguez2012,QuinteroNoda2016}.
Alternative spectral shapes, such as the Gaussian and Lorentzian profiles, are useful when trying to represent profile shapes manifesting via different radiative transfer effects that are often present in the spectra.

Previous studies \cite{Grant2018} have attempted to fit complex {\CaIR} line profiles using a linear combination of two Gaussians, one for each component of the atmosphere attempting to be modelled. 
Boundary conditions for each fitted profile are employed such that the first Gaussian models an absorption component representative of the quiescent atmosphere, while the second Gaussian models an emission component linked to the dynamic phenomenon under investigation (e.g., magnetoacoustic shocks).
This works well when two distinct components are present in an observed line profile, yet it introduces a number of challenges when only a single absorption component is resolved.
Firstly, it takes more computational time and resources to fit the two-profile combination than a single profile, since twice the number of fitting parameters need to be constrained.
Secondly, the boundary conditions placed on the fitted Gaussians tend to be subjectively chosen, and therefore may subject any extracted results to a user bias. 
Thirdly, the most problematic issue is the tendency for the two-profile combination to overfit the observed data as the fitting algorithm naturally attempts to smooth out noise and asymmetries in the observed spectra.
This results in neither of the constituent profiles, by themselves, accurately modelling the absorption components present in the spectra, and Doppler shifts extracted from any one of the two fitted components would be potentially misleading and incorrect. 

A novel method to alleviate the degree of human interaction required when processing large datasets hinges on the concept of machine learning, which is becoming increasingly commonplace in the field of solar physics.
Machine learning, in particular neural networks, have been used in solar physics research from at least the early 1990s for applications such as predicting the maximum number of sunspots during a solar cycle \cite{Koons1990}, and also predicting the number of sunspots produced throughout the Sun's lifetime \cite{Calvo1995}.
More recently, they have been used for predicting and detecting features on the Sun such as flares \cite{Qahwaji2007}, coronal mass ejections \cite{Sudar2016}, and active regions on the far side of the solar surface \cite{Felipe2019}.
On small spatial scales, machine learning has been shown to play an important role in the investigation of velocity flows \cite{AsensioRamos2017} and rapid spectropolarimetric inversions of Stokes~$I/Q/U/V$ spectra \cite{AsensioRamos2019}. 
Hence, the application of machine learning and neural networks to the challenging problem of fitting complex spectral line profiles (e.g., {\CaIR}) is both timely and important for the accurate study of small-scale dynamic phenomena that are ubiquitously observed to permeate the solar atmosphere.

In this paper we detail a method for fitting profiles to spectral lines that often contain multiple atmospheric components.
Our method uses machine learning techniques to distinguish between spectra that have multi-component atmospheres and those that can be best represented by single spectral fits, allowing us to adjust the model and fitting method accordingly.
We also provide a proof-of-concept study on a challenging {\CaIR} spectral imaging dataset to show the suitability of this technique for widespread usage.

\section{Observations}
\label{sec:observations}

\begin{figure}[!t]
	\centering
	\includegraphics[width=\textwidth]{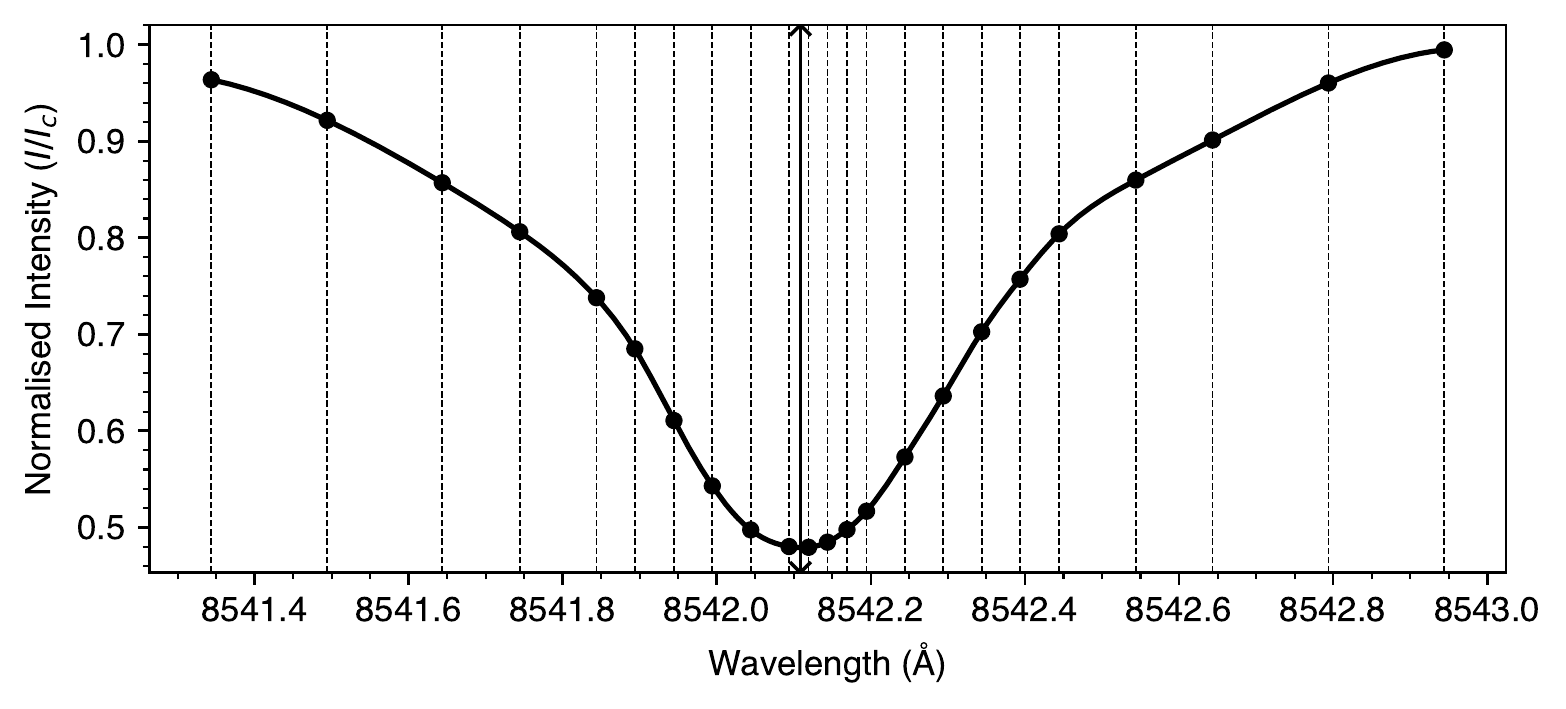}
	\caption{Plot of the average {\CaIR} spectrum taken from the quiet Sun region shown in Figure~\ref{fig:coaligned}, where the intensities, $I$, have been normalised by the quiet Sun continuum intensity, $I_{c}$. Solid black markers and vertical dashed lines highlight the 23 wavelength points used in our analysis out of the 27 wavelength points that were sampled by IBIS. The vertical solid line, emphasised by arrows, represents the stationary line core wavelength. Each spectral scan took 5.8~s to complete.}
	\label{fig:averagespectrum}
\end{figure}

Spectral imaging observations of active region NOAA 12149 were acquired from 14:37 -- 17:37~UT on 2014 August 30 using the Interferometric BIdimensional Spectrometer (IBIS;\cite{Cavallini2006}) instrument at the National Science Foundation's Dunn Solar Telescope, New Mexico, USA. 
IBIS is a Fabry-P{\'{e}}rot instrument able to obtain high temporal, spatial, and spectral resolution imaging spectroscopy measurements of the solar atmosphere.
The {\CaIR} spectral line was chosen for our observations, with IBIS operating in a Stokes~$I$ imaging mode (i.e., no polarimetric measurements were acquired) in order to maximise the temporal resolution of the spectroscopic scans.  
The observations obtained consisted of 2103~spectral scans utilising 27 non-equidistant wavelength points sampling over a 2.4{\,}{\AA} window centred on the {\CaIR} line core. However, for the purposes of subsequent analysis, we focus on the central 23 wavelength points over a 1.6{\,}{\AA} window centred on the {\CaIR} line core. As shown in Figure~\ref{fig:averagespectrum}, closer wavelength spacing was chosen around the line core to provide better quiescent profile fitting.
The spatial sampling was $0{\,}.{\!\!}{''}098$ per pixel, and the cadence of each {\CaIR} scan was $5.8$~s.

A contextual continuum image of active region NOAA~12149 was acquired from the Helioseismic and Magnetic Imager (HMI;\cite{Schou2011}) onboard the Solar Dynamics Observatory (SDO;\cite{Pesnell2011}).
This measurement was taken at the start of the IBIS spectral imaging sequence, and once processed through standard SunPy data reduction algorithms \cite{Barnes2020,Mumford2020} provided a full disk reference image with a spatial sampling equal to $0{\,}.{\!\!}{''}6$ per pixel.  
Using the HMI contextual image to define absolute solar coordinates, the IBIS dataset was subsequently co-aligned to it using cross-correlation techniques \cite{Jess2012}.
Figure~\ref{fig:coaligned} shows example images from the IBIS dataset co-aligned with the SDO/HMI continuum.

\section{Methods}
\label{sec:methods}
Presented in this section are the details of a method which utilises a neural network to classify a spectrum based on its profile shape.
A suitable spectral fitting model is then selected for the spectrum according to its neural network classification.
By accurate fitting of the chosen model, a Doppler shift, and therefore a velocity, can be uncovered for each of the constituent spectral components, thus allowing quiescent and dynamic parts of the atmosphere to be isolated and studied independently.

\begin{figure}[!t]
	\centering
	\includegraphics[width=\textwidth]{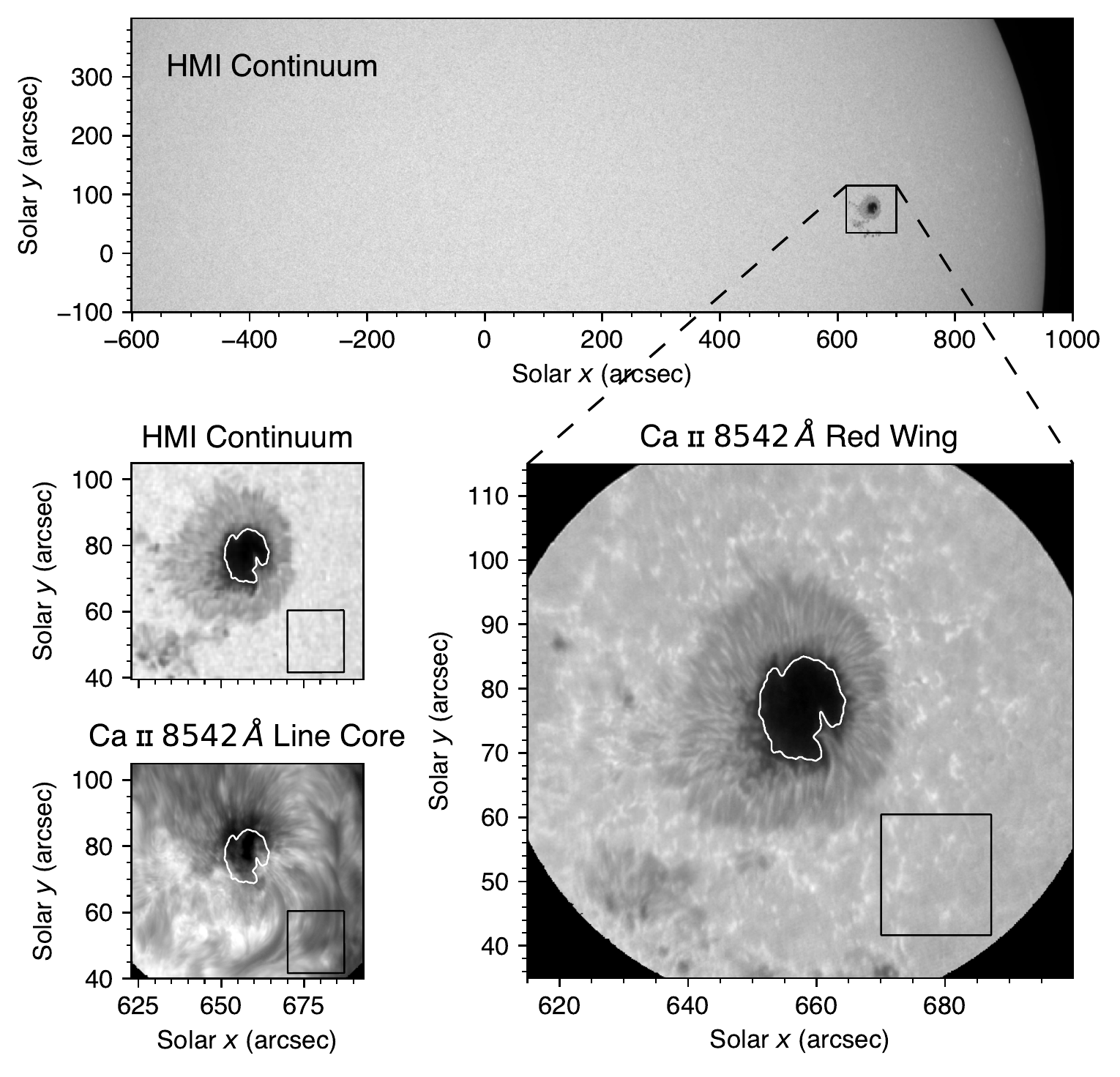}
	\caption{Co-aligned images of the IBIS {\CaIR} red wing (line core $+1.2$~{\AA}; lower right), IBIS {\CaIR} line core (lower left), and SDO/HMI 6173{\,}{\AA} continuum (middle left) intensities. The location on the solar disk of active region NOAA~12149 during the observation period is shown in the top panel using a solid black box. The solid white contour used in each of the images represents the umbra-penumbra boundary used to isolate the umbral regions. The black rectangle displayed in each of the lower panels encloses a quiescent region of the solar atmosphere used to determine the average quiet Sun profile discussed in Section~\ref{sec:methods}.}
	\label{fig:coaligned}
\end{figure}

An average quiet Sun spectrum was extracted from the dataset by averaging all of the IBIS spectra contained within the rectangular region plotted in Figure~\ref{fig:coaligned} across all 2103~scans. 
The average quiet Sun spectrum incorporated the averaging of 70{\,}660{\,}800 individual spectra, with the resulting profile plotted in Figure~\ref{fig:averagespectrum}.
This location was chosen for constructing the quiet Sun spectrum because it is isolated from both the sunspot umbra (where complex {\CaIR} profiles synonymous with shock formation are known to form) and the regions containing magnetic pores and bright points that often result in enhanced {\CaIR} line core emission due to the formation of chromospheric plage \cite{Pietarila2007a,Pietarila2007b,delaCruzRodriguez2013a}. 
The quiet Sun spectrum shown in Figure~\ref{fig:averagespectrum} demonstrates a deep absorption profile with no clear evidence of enhanced wing or line-core emission, and hence forms a good `rest' profile of the solar atmosphere that can be used to benchmark spectral fluctuations resulting from dynamic {\CaIR} phenomena. 

\subsection{Choosing a fitting profile}
\label{sec:fittingprofile}
Observations of a spectral line corresponding to a specific atomic electron transition provides emission and absorption signatures over a range of wavelengths (i.e., not just an infinitely narrow feature at the wavelength associated with the transitioning electron).
This increased width of the spectral line is due to a number of effects including Doppler broadening, pressure broadening, and Zeeman splitting \cite{Shine1972,Cauzzi2008,delaCruzRodriguez2012,QuinteroNoda2016}.  
These effects can be replicated closely by convolving a Gaussian function with a Lorentzian profile, producing a Voigt function, $V$, defined by \cite{Zaghloul2007},  
\begin{align}\label{eq:voigt}
V(x;A, \sigma, \gamma) = A \int_{-\infty}^{\infty} G(u;\sigma) L(x-u;\gamma) \mathrm{d}u \ ,
\end{align}
where $A$ is a constant scaling for the amplitude, $G(x; \sigma) = \exp(-x^2/(2\sigma^2)) / (\sigma\sqrt{2\pi})$ is a Gaussian function centred at zero, and $L(x; \gamma) = \gamma / (\pi (x^2 + \gamma^2))$ is a Lorentzian function centred at zero. The parameters $\sigma$ and $\gamma$ are the standard deviations of the Gaussian and the half-width at half-maximum of the Lorentzian, respectively.
The variable of integration is $u$.
Since the Lorentzian and Gaussian functions are centred on zero, the variable $x$ that is passed into the Voigt function must first be shifted by subtracting the wavelength of the line core, $x_0$, such that the line core wavelength is at zero in the corresponding function. 
The $x_0$ parameter must be re-added to the fitted spectra in order to precisely account for Doppler-shifted plasma.

A linear combination of Voigt profiles was chosen to model the absorption and emission components of the spectra, hence producing a `double Voigt' model.
If the chosen spectrum shows only emission or absorption profile shapes (e.g., like shown in Figure~{\ref{fig:averagespectrum}}), then a `single Voigt' model is selected, thus avoiding the computational and overfitment drawbacks associated with multiple component fits when not necessary. 
Note that a background intensity level is not included in the generated models computed from Equation~{\ref{eq:voigt}} as this will be calculated and subtracted from the spectra before fitting, as detailed in Section~3d(\ref{sec:backgroundremoval}).
By modelling the spectra with Voigt profiles instead of just simple Gaussian functions, we can fit the spectral lines more accurately as we are also including line wing broadening that mimics the observed spectral lines. 

Single or double Voigt profiles are fitted iteratively to each spectrum until it is sufficiently similar to the observed line profile.
During the fitting process, the employed model (`single Voigt' or `double Voigt') is statistically evaluated numerous times to monitor the level of convergence between the synthetic and observed line profiles. 
It is therefore necessary for the selected model to be able to be evaluated efficiently, thereby saving computational time.
The Voigt function was implemented in Python with the convolution being carried out by a numerical integration function, \texttt{quad}, provided by the \texttt{scipy.integrate} Python module\cite{Virtanen2020,Virtanen2019}.
An approximation of the Voigt profile\cite{McLean1994} was also explored. However, even though this function requires less time to compute when compared to the numerical integration approach, the fitting methods exhibited very slow convergence, and hence we adopted the numerical integration technique for the remainder of our work.

We utilise a number of techniques to maximise the efficiency of our calculated Voigt functions.
The integrand of the Voigt function is written in the C programming language and compiled into a shared library.
This C function can then be imported into Python using the \texttt{ctypes} library.
The \texttt{scipy.integrate} numerical integration function is then able to use this more efficient function as an integrand.
We also adjust the absolute and relative error tolerances for the numerical integration, setting them to 0.149 and 0.000149, respectively. 
In practical terms, this means that the error in the raw intensity value at each wavelength point will be $0.149$ for $|I-I_{\text{BW}}| \leq 1000$ and $0.000149 \times |I-I_{\text{BW}}|$ for $|I-I_{\text{BW}}| > 1000$, where $I$ are the intensity values at a each wavelength point on a profile, and $I_{\text{BW}}$ is the blue wing (rest line core $- 1.2$~{\AA}) intensity value for the same profile.

Taking a typical full IBIS imaging spectral scan from our dataset, we computed the $|I-I_{\text{BW}}|$ value at every intensity value across all 664{\,}796 individual spectra in the scan.
We then calculated the error in each of the raw intensity values, $I$, using these computed $|I-I_{\text{BW}}|$ values.
By dividing each error by $I$, a percentage error was determined for every intensity value.
Over the entire field-of-view, the average percentage error was 0.00666\%, with an error of 0.00739\% over the subset of wavelength points $\pm 0.1$~{\AA} around the stationary line core. 
For the umbral spectra, which have much lower $I_{\text{BW}}$ values, the average percentage error was 0.0179\%, with an error of 0.0180\% around the stationary line core. 
In our particular test, the maximum percentage error across the whole scan was only 0.0488\%. 
Given that the IBIS spectra typically have line core and wing intensity values exceeding 1730 and 2550 detector counts, respectively, and $|I-I_{\text{BW}}|$ values of 757 and 149, respectively, these values correspond to absolute intensity errors of $0.149$~detector counts. As such, these small percentage error tolerances are sufficient for our investigation.

\subsection{Neural network}
\label{sec:neural}
\subsubsection{Classifying spectra}
\label{sec:classifying}
As discussed in the Section~{\ref{sec:introduction}}, fitting a double Voigt model to all spectra introduces temporal efficiency and overfitting problems for those that contain only a single absorption/emission component.
In these cases, a single Voigt model would be more appropriate.
Our method for fitting spectral lines employs machine learning techniques to tailor the specific model (i.e., single Voigt or double Voigt functions) used for each spectrum sampled.
This way, if a spectrum only contains a single component atmosphere consisting of an absorption dip or an emission peak, a single Voigt profile is fitted to the spectrum, with boundary constraints chosen objectively for such a line profile.
On the other hand, if two components are detected, one absorption and one emission component, then a double Voigt profile is fitted, with each Voigt function appropriately constrained to fit the absorption and emission components of the observed line profile.
In order to perform this automated task, we require a robust methodology to be able to reliably identify how many spectral components are present in each observed spectrum.

Machine learning, specifically artificial neural networks, are used to classify the spectra based on the amount of emission relative to absorption present in profile (see Section~3{\ref{sec:training}} for an in-depth discussion of how this is performed).
Due to the diversity of spectral profiles in our dataset, it would have been very challenging to devise a fixed criterion that can be implemented into an algorithm and used to produce classifications more accurately than our neural network classifier.
The neural network was trained to assign one of five classifications to each observed spectrum (see Figure~\ref{fig:classifications}), with the assigned classification used to adjust the model employed for final fitting.
Specifically, the five classifications chosen consist of, 
\begin{enumerate}[label=\arabic*~---]
\addtocounter{enumi}{-1}
    \item A profile exhibiting purely absorption characteristics, with no evidence of emission or asymmetric line wings;
    \item An absorption line profile similar to the classification `0', but with evidence of a less precise line profile minimum due to, e.g., flattening of the line core intensities;
    \item A profile that begins to show signs of embedded emission characteristics, yet the line core intensity is still less than the continuum intensity;
    \item A line profile similar to classification `2', but with enhanced emission signatures such that the line core intensity is now just above the continuum level; and
    \item A line profile showing heightened emission that is considerably brighter than the neighbouring continuum, hence dominating the resultant spectral profile.
\end{enumerate}

This classification regime was chosen as it allows key features of the spectral shape to be identified and used to adapt the model, and it also identifies regions where the level of emission is particularly high.
The neural network infrastructure is provided by the {\it scikit-learn} Python module \cite{Pedregosa2011,Buitinck2013,Grisel2020}.
From this package we use the multi-layer perceptron (MLP) classifier\cite{Costa1994} along with the L-BFGS-B solver\cite{Byrd1996}, which is a unique version of the Broyden-Fletcher-Goldfarb-Shanno (BFGS) solver for limited memory and boundary constrained problems.

\subsubsection{Mathematical processes}
\label{sec:mathematics}

The observed spectrum is first interpolated on to a constant wavelength grid, resulting in a profile consisting of 33 data points.
Our neural network implementation, with the multi-layer perceptron (MLP) classifier, takes the interpolated spectrum as a one-dimensional input layer with 33 neurons.
The neural network then standardises its input by rescaling the input vector to range from 0 to 1.
As a general overview, each of the neurons of the input layer are subsequently connected to each of forty neurons on a single hidden layer.
Finally, each neuron of the hidden layer is connected to all of the classifications on the five neuron output layer.
The classification with the largest probability is assigned to the specific spectrum.

More specifically, each neuron-neuron connection between two layers has a weight, $w_{ij}$, where $i$ is the index of the neuron in the current layer, and $j$ is the index of the neuron on the previous layer.
All of the neurons of the hidden and output layers have a bias, $b$.
The input is propagated through the neural network along the connections between neurons of adjacent layers, with the applied weights and biases determining the level of connectivity between the subsequent layers.
The result at each neuron is calculated using the equation,
\begin{align}
y = g\left( \sum_j w_j x_j - b \right) \ ,
\end{align}
where $x_j$ are the outputs from previous layers, $w_j$ are their associated weights, and $g$ represents the activation function.

The activation function for the neurons of the hidden layer was chosen to be the rectified linear unit function (ReLU; \cite{Hahnloser2000}) which is defined as $f(x)=\max(0,x)$, meaning that the neuron only `activates' and passes a non-zero value on to the next layer if its result is positive.
The output layer employed the softmax activation function, which is defined as $\text{softmax}(x)_i = \exp(x_i)/(\sum_{k=1}^5 \exp(x_k))$, where $i \in \{1, 2, 3, 4, 5\}$ are the five possible classifications that the neural network can assign.

This equation can be extended to apply to the output of a whole layer.
Firstly, the outputs from the previous layer, $x_j$, can be represented as a column vector, $\bm{x} \in \mathbb{R}^n$, where $n$ is the number of neurons in the previous layer.
Secondly, the weights, $w_{ij}$, can be represented as a matrix, $\bm{W} \in \mathbb{R}^{m \times n}$, where $m$ is the number of neurons in the current layer.
Importantly, the matrix of weights between the input layer and the one hidden layer is given by $\bm{W}^{(01)}$, and the input values is given by $\bm{x}^{(0)}$.
Similarly, $\bm{W}^{(12)}$ represents the matrix of weights between the hidden layer and output layer, with $\bm{x}^{(1)}$ representing the output of the hidden layer.
The biases, $\bm{b} \in \mathbb{R}^m$, for each neuron of the hidden layer is given by $\bm{b}^{(1)}$, while the output layer can be represented by $\bm{b}^{(2)}$.
The output of the neural network is therefore $\bm{y} \in \mathbb{R}^5$.
As a result, a spectrum can be classified by evaluating the following set of equations in order,
\begin{align}
	\bm{x}^{(1)} &= f\left(\bm{W}^{(01)}\bm{x}^{(0)} + \bm{b}^{(1)} \right) \\
	\bm{y} &= \text{softmax}\left(\bm{W}^{(12)}\bm{x}^{(1)} + \bm{b}^{(2)} \right)
\end{align}
where $f : \mathbb{R} \rightarrow \mathbb{R}$ and $\text{softmax} : \mathbb{R} \rightarrow \mathbb{R}$ are the ReLU and softmax activation functions, respectively, applied to the matrix elements.
Subsequently, the spectrum is then assigned a classification associated with the largest output neuron value.
Many spectra can be classified simultaneously by increasing the dimensions of the matrices forming tensors, which can be operated very efficiently using graphics processing units (GPUs; \cite{Bergstra2010}).

The architecture of the neural network described in this study can classify a full IBIS imaging spectral scan, consisting of 664{\,}796 individual spectra contained within the circular aperture of the instrument (see Figure~{\ref{fig:coaligned}}), in approximately 48 seconds using a single 2.10~GHz Intel Xeon CPU core. 
As a result of this reasonable timeframe, porting to GPUs was not investigated in the present study. 
However, with larger field-of-view sizes and more rapid cadences from upcoming next-generation instrumentation \cite{Kentischer2012}, the resulting increase in data volume may require attention to be turned towards GPUs to provide the necessary accelerated performance to implement this technique in near real-time on next-generation datasets.

\subsection{Training procedure}
\label{sec:training}
In Section~3{\ref{sec:neural}}, the weights and biases are initially unknown.
For the neural network to be able to accurately classify spectra, the optimal weights and biases must be calculated, which is achieved by fitting the parameters of the neural network to a set of manually labelled training spectra.

The current accuracy of the neural network is quantified by a loss function, calculated using the manually assigned classification and the classification currently being assigned by the neural network.
The weights and biases are optimised by minimising the value of the loss function.
In this method we use the log-loss function,
\begin{align}
\text{log-loss} = \dfrac{1}{N_S} \sum_{(y, y') \in S} \left( - y \log(y') - (1 - y) \log(1 - y')\right) \ ,
\end{align}
where $S$ is the set of ground truth labels ($y$), defined as either `0' or `1', and currently predicted probabilities (\(y'\)) spanning a range between 0 and 1, for each classification of the spectra in the training set.
The parameter $N_S$ is the number of $(y, y')$ pairs in the set $S$.

To find the predicted probabilities for each classification, the algorithm passes the training set of spectra through the neural network, configured with an initially random set of weights and biases, before calculating the subsequent log-loss function.
Finally, the L2 regularisation term,
\begin{align}
\text{L2} = \frac{\alpha}{2N_S} \left( ||W^{(01)}||^2_2 + ||W^{(12)}||^2_2 \right) \ ,
\end{align}
where $||W||^2_2$ is the sum of the weight matrices squared and $\alpha$ is the parameter that controls the amount of regularisation, is added to the loss function. 
The training algorithm iteratively tries a number of different $\alpha$ terms, before selecting the specific value that produces the most accurate classifications after training.

The weights and biases that minimise the loss function are found using the L-BFGS-B quasi-Newton minimisation algorithm\cite{Byrd1996}, a variant of L-BFGS.
This efficient algorithm works by finding an approximation of the Hessian matrix, which is a matrix of second derivatives of the loss function with respect to the weights and biases.
Importantly, the Hessian matrix forms part of the quadratic term in the second-order Taylor expansion of the loss function when evaluated for a one-dimensional vector composed of the current weights and biases.
The inverse of the Hessian matrix can also be found efficiently, which is then used to determine the step direction the vector of weights and biases should take for minimisation of the loss function.

There are two separate sets of manually labelled ground truth data, one used for training the neural network and another used for testing its performance.
By using a different set of spectra to test the performance of the neural network, we are able to detect if the neural network is overfitting, which would result in the test set being assigned less accurate classifications. 
Various statistics are determined, including a calculation of the percentage of spectra the neural network correctly assigned the ground truth classification.

The MLP classifier in the {\it scikit-learn} module also has a number of parameters that must be specified before the training is carried out.
These include the numbers of hidden layers and neurons employed, the $\alpha$ parameter used for L2 regularisation, the activation function used for the hidden layers (ReLU), and the algorithm used to optimise the weights and biases. 
A neural network consisting of a single hidden layer with forty neurons was chosen for our study.

The L-BFGS, stochastic gradient descent (SGD), and Adam \cite{Kingma2014} optimisation algorithms were trialled.
Upon successive trainings, the L-BFGS method produced trained neural networks with the most consistent accuracy score, and hence this approach was adopted for the current study.
The accuracy score is defined as the fraction of the testing set of spectra that the neural network successfully assigned its manually labelled classification.
Our method uses the \texttt{GridSearchCV} infrastructure provided by the {\it scikit-learn} module to select the best \(\alpha\) parameter from the list \([1,2,3,4,5,6,7,8,9] \times 10^{-5}\).

\subsubsection{Training with the IBIS dataset}
\label{sec:IBIStraining}
\begin{figure}[!t]
	\includegraphics[width=\textwidth]{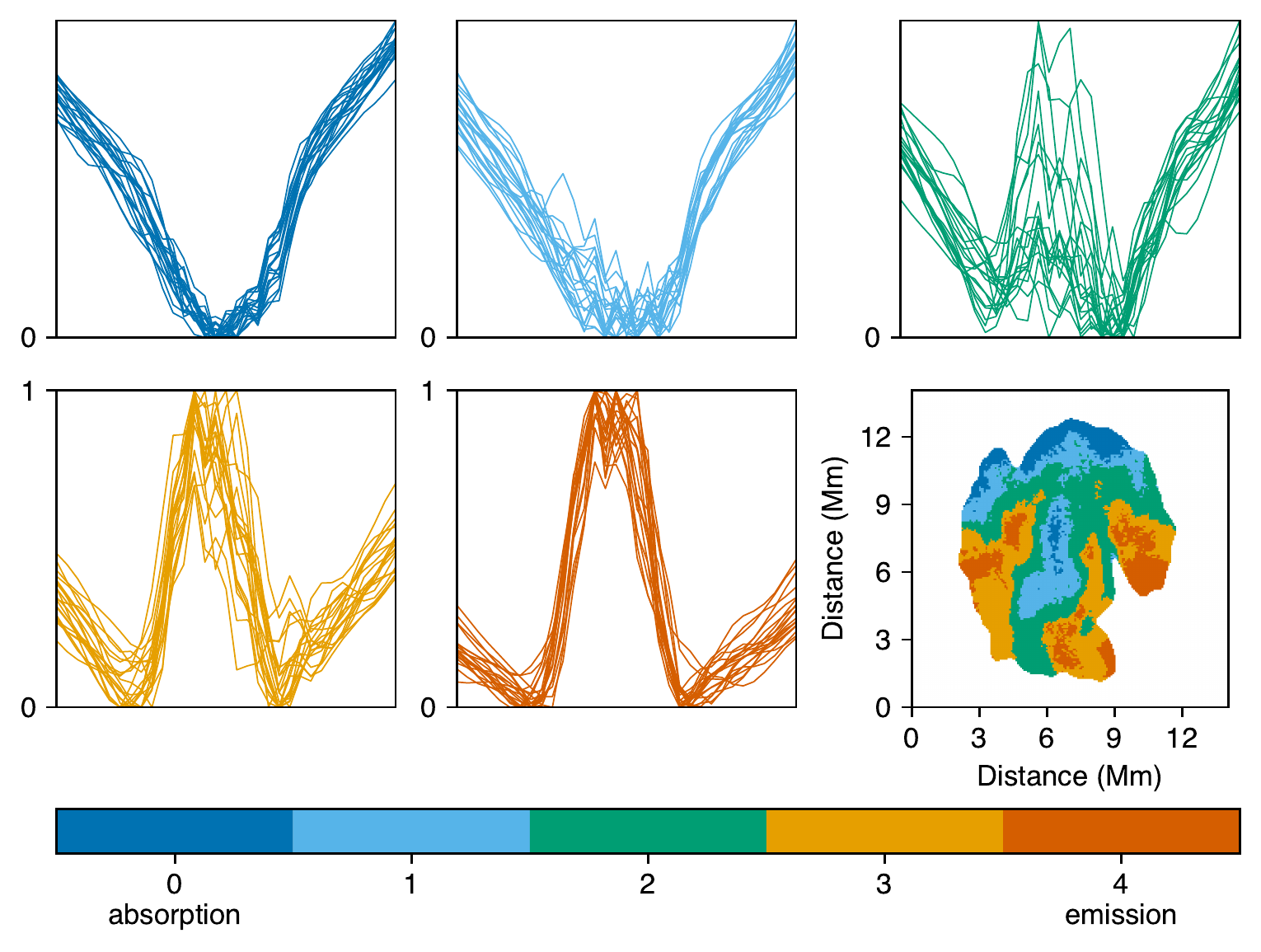}
	\caption{Plots of stacked {\CaIR} umbral line spectra grouped by their neural network classification, where the intensity scale for each spectrum is normalised between 0 and 1 to aid visualisation. A two-dimensional map (lower right) reveals the neural network classifications for the {\CaIR} spectra present within the umbra (see contours in Figure~{\ref{fig:coaligned}}) for a single IBIS spectral imaging scan.} 
	\label{fig:classifications}
\end{figure}

To train and test the neural network, 200 spectra from the IBIS {\CaIR} dataset introduced in Section~{\ref{sec:observations}} were chosen at random from within the umbra.
The spectra in this set are manually labelled with a classification, providing ground truth data for training and testing purposes.
The first half of this 200 spectra dataset is used to train the neural network, while the other half is used to test the success of the trained neural network.
A subset of these spectra are separated into their manually labelled classifications and plotted in Figure~\ref{fig:classifications} for visual clarity.
As described in Section~3b({\ref{sec:classifying}}), these spectra are classified based on the ratio of emission to absorption, with the extreme ends of the scale being those spectra showing no evidence of emission (classified as `0'), and those spectra with obvious emission much brighter than the neighbouring continuum (classified as `4').

Following the methodology outlined in Section~3{\ref{sec:training}}, the success of the trained neural network can be classified in terms of `precision' and `recall' statistics.
Here, the precision is defined as the percentage of spectra that were correctly assigned a classification, $c$, from all the spectra that were also assigned the same classification by the neural network.
Recall is defined as the percentage of spectra that were correctly assigned a classification, $c$, out of all the spectra that should have been assigned this specific classification according the the ground truth data.
For each of the classifications (0 $\rightarrow$ 4) in ascending order, the precisions obtained were 77\%, 74\%, 71\%, 100\%, and 94\%.
Similarly, in terms of recall, the percentages obtained were 94\%, 78\%, 80\%, 62\%, and 100\%.
These result in overall averages, weighted by the number of ground truth spectra per classification, of 83\% and 81\% for the precision and recall parameters, respectively.

Importantly, 95\% of the spectra incorrectly classified by the neural network had a deviation of only one classification number.
These performance measures suggest that the neural network is sufficiently accurate for our purposes.
Running the neural network classification algorithm on the umbral pixels extracted from a single IBIS spectral imaging scan displays the expected complexities associated with this type of solar feature. In particular, the lower right panel of Figure~\ref{fig:classifications} displays the classifications that the trained neural network returned, revealing a multitude of purely absorption (i.e., quiescent) and emission (i.e., active) spectra.

\subsection{Fitting method}

The following preprocessing steps and fitting methods are applied to each spectrum of the IBIS dataset independently.
Based on the neural network processing described in Section~3c({\ref{sec:IBIStraining}}), each spectrum is assigned a classification. 
Spectra that are assigned a classification of either `0' or `1' are fitted with a simple, single Voigt profile as described in Section~3{\ref{sec:fittingprofile}}.
Furthermore, spectra assigned a classification of either `2', `3', or `4' are modelled with double Voigt profiles.
This results in spectra with a dominant absorption component being fitted with a single quiescent atmosphere model, whereas spectra with increasing emission components are modelled with multi-component fits that are more representative of the dynamic atmosphere.
Doppler velocity information can then be extracted for both the quiescent and active atmospheres using the parameters inferred from the fitted models.

As will be discussed in Section~3d(\ref{sec:weightedfit}), we will further distinguish between classifications `0' and 1 $\rightarrow$ 4. However, classifications 2 $\rightarrow$ 4 are processed in an identical manner. While we found no additional benefits to distinguishing between these different classifications when fitting, we have kept these as distinct classifications to demonstrate the adaptability of our method and to assist with relating the spatial distribution of classifications to any dynamic features found in the Doppler velocity measurements.

\subsubsection{Preprocessing techniques}

As shown in Figure~\ref{fig:averagespectrum}, the spectral images acquired using IBIS consist of intensity measurements obtained across 27 non-equidistant wavelength points.
However, in our analysis we focus on the central 23 wavelength points, providing a $1.6$~{\AA} wavelength range across the Ca~{\sc{ii}} line, allowing Doppler velocities spanning $\approx\pm30$~km/s to be studied.
Each spectrum is interpolated on to an equidistant wavelength grid employing a wavelength spacing of 0.05\,{\AA}. 
This produced a spectrum comprising of 33 equidistant wavelength samples.

The next stage is to ensure the line wings are correctly calibrated.
IBIS includes a prefilter with a FWHM of $4.6${\,}{\AA} centred at $8542.5${\,}{\AA} in order to isolate the 8542{\,}{\AA} signal from the other resonant wavelengths produced by the Fabry-P{\'{e}}rot interferometer \cite{Cauzzi2008}.
As a result, each profile is initially corrected by removing the prefilter transmission response from the measured spectrum.

\subsubsection{Background removal}
\label{sec:backgroundremoval}

In the fitted models, a background profile is not included, and as a result the continuum intensity is expected to be zero.
The motivation for not including a background profile is to reduce the number of free parameters, and thus make the model less computationally challenging to fit.
Hence, a constant background intensity is calculated for each spectrum by performing a boxcar moving average of the intensities calculated 1.2{\,}{\AA} into the blue wing from the {\CaIR} line core over a time frame of $\pm 2$~minutes relative to the current scan, which is subtracted from the spectrum before fitting with a suitable model. 
When processing a number of spectra over a large field-of-view, the calculation of the backgrounds and their subsequent subtraction can be vectorised, resulting in a significant computational speed improvement compared to processing each spectrum individually.
Due to the breadth of the {\CaIR} line, in conjunction with the relatively narrow wavelength range of the IBIS prefilter, it is not possible to to calculate the true continuum intensity for our dataset. As such, we approximate the continuum intensity, $I_{c}$, using a boxcar moving average placed at the furthest blue-wing position of the IBIS spectral imaging scans (i.e. line core $- 1.2${\,}{\AA}).

\begin{figure}[!t]
	\centering
	\includegraphics[width=0.8\textwidth]{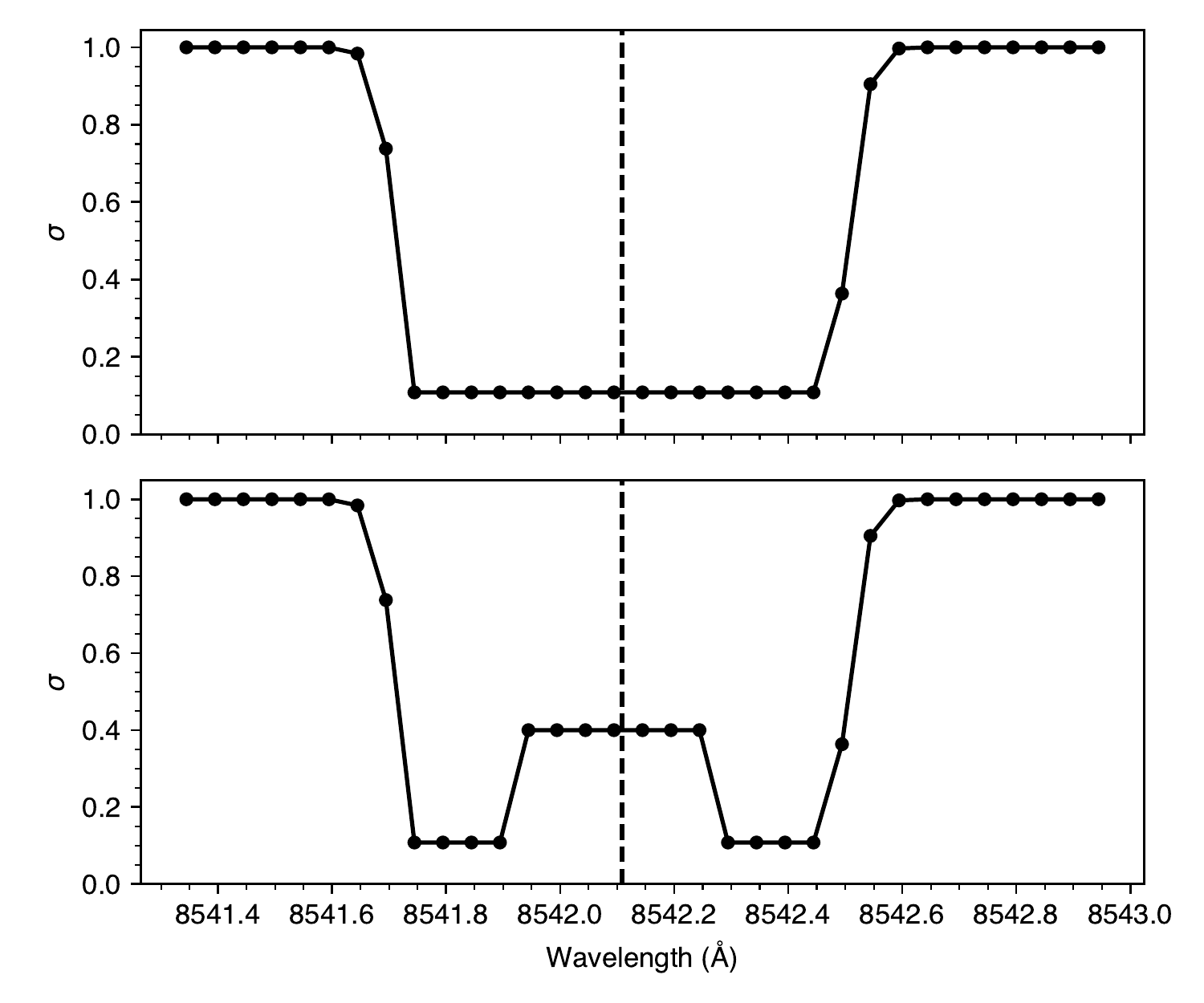}
	\caption{Plots of the sigma profiles used to weight particular regions of the spectra during the fitting process. The top profile is used for spectra of classification `0' and has the effect of reducing the priority of fitting the wings. The bottom profile is used for all other classifications of spectra and reduces the priority of fitting the precise shape of the spectral turning point. The vertical dashed line represents the stationary line core wavelength that these profiles are mapped on to.}
	\label{fig:sigma}
\end{figure}

\subsubsection{Weighted fit}
\label{sec:weightedfit}

In order to fit the chosen model to the spectra more accurately, we applied weights to different parts of the spectrum.
Weights are given by assigning error bars to the spectrum, which the fitting algorithm uses when calculating goodness-of-fit parameters.
As the measurement of Doppler velocities is of paramount importance, especially for quiescent spectra classified as `0' where the Doppler shifts may be subtle, we prioritise fitting around the spectral line core.
As a result, we add larger uncertainties to the wings of the spectral line when compared to the line core, which can be visualised in the upper panel of Figure~{\ref{fig:sigma}}.

On the other hand, for spectral classifications 1 $\rightarrow$ 4, we found that the accuracy of the fit was improved by increasing the associated uncertainties at wavelengths immediately surrounding the wavelength of the stationary line core (see the lower panel of Figure~{\ref{fig:sigma}}).
This had the effect of lowering the priority for fitting the exact shape of the peak, which was either difficult to isolate (e.g., for classifications `1' and `2' shown in Figure~{\ref{fig:classifications}}), or incentivised the optimisation algorithms to fit more closely the steepening gradients of the spectral line wings (e.g., for classifications `3' and `4' shown in Figure~{\ref{fig:classifications}}) that are representative of the dynamic activity present in the spectra.

\begin{figure}[!t]
	\centering
	\includegraphics[width=0.6\textwidth]{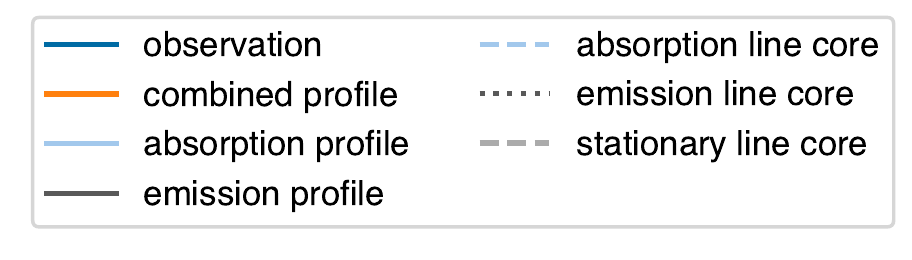}
	\includegraphics[width=0.49\textwidth]{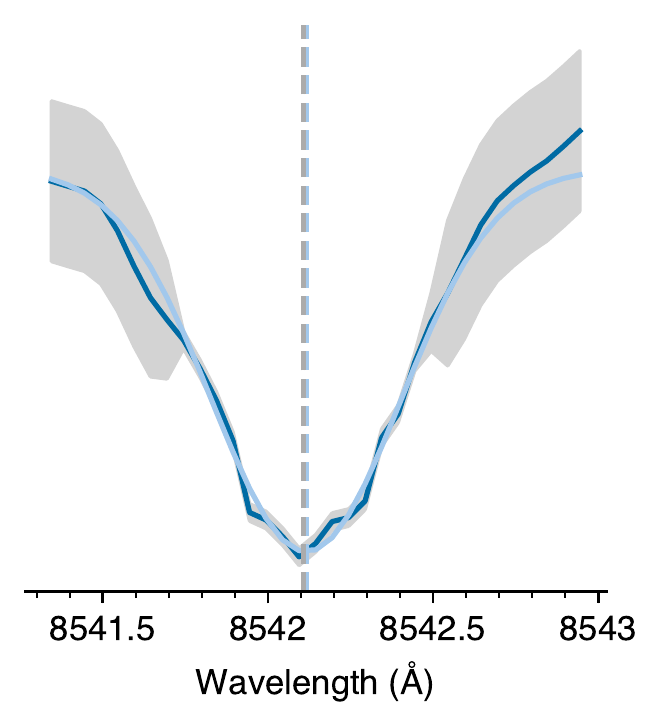}
	\includegraphics[width=0.49\textwidth]{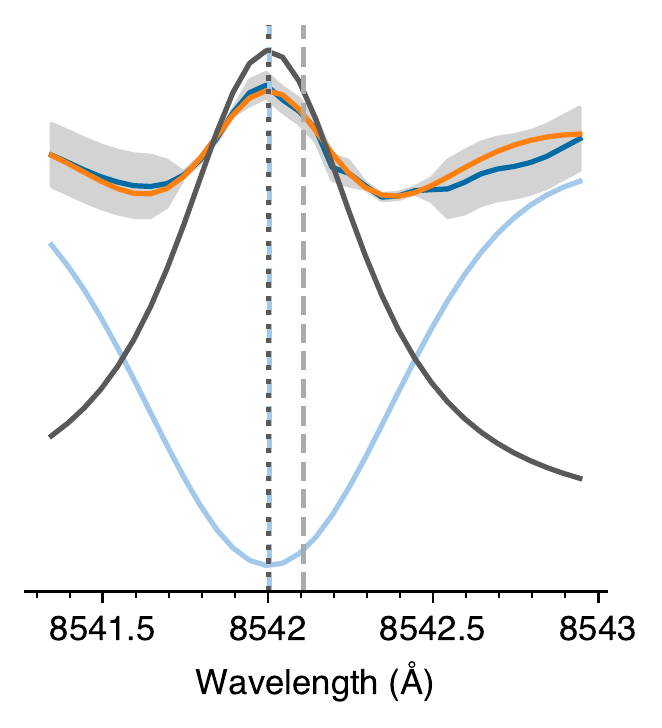}
	\caption{Examples showing the fitting methods applied to spectra with a single component atmosphere (neural network classification `0'; left panel) and a multi-component atmosphere (neural network classification `3'; right panel). The labelled vertical lines identify the line core wavelengths of each atmospheric component fitted, i.e., the central wavelengths of the relevant Voigt functions. The addition of the multi-component fits (right panel) shows close agreement with the observed profile. The grey shaded regions plotted on top of the spectra represent the sigma weighting profiles displayed in Figure~{\ref{fig:sigma}}.}
	\label{fig:fits}
\end{figure}

\subsubsection{Least squares optimisation}
The chosen model (i.e., either a single or double Voigt model) is fitted to the spectra using the \texttt{curve\_fit} function provided by the \texttt{scipy.optimize} Python module\cite{Virtanen2020,Virtanen2019}.
With this function we use the Trust Region Reflective algorithm for least squares optimisation, similar to that proposed by Branch et~al.\cite{Branch1999}, which allows for bounds to be specified for the parameters intrinsic to the model, which assists with spectral convergence.

The amplitude is bounded such that a single Voigt model must fit a negative (i.e., absorption) amplitude, while a double Voigt model must fit one negative amplitude (representing the quiescent atmosphere) and one positive amplitude (representing the active atmospheric component).
This helps to prevent overfitting issues described in Section~{\ref{sec:introduction}}.

The wavelength at the centre of the absorption Voigt profile is constrained such that it does not deviate too far from the stationary line core wavelength.
The central wavelength of the absorption Voigt function is allowed to vary within a {0.15\,\AA} window above and below the stationary line core wavelength.
The stationary line core wavelength is found by applying this method, without bounds, to the average quiet Sun spectrum introduced in Section~\ref{sec:observations}.
This constrains the method to only allow Doppler velocities in the absorption component that are \(\pm 5.27\)\,km/s.
This helps ensure that the model represents what is happening physically in the spectra, and does not give an unphysical Doppler shift. 

Bounds are also placed on the $\gamma$ and $\sigma$ parameters for the Lorentzian and Gaussian shapes that are convolved to form the resultant Voigt profile. 
The bounds chosen for the $\gamma$ and $\sigma$ components are $10^{-6} < \gamma~\text{or}~\sigma < 1$, which allows a wide range of spectral broadening to be accounted for.
An initial guess is also supplied to the fitting function, where the central wavelength of each constituent Voigt profile is initially set to the stationary line core wavelength.
Any Voigt profiles modelling absorption are given a negative initial amplitude that is typical for the spectra dataset, while any Voigt profiles representing emission are given a typical positive initial amplitude. 
All $\gamma$ and $\sigma$ initial guesses are taken to be $0.1$ and $0.2$, respectively.

Figure~\ref{fig:fits} shows examples of the fitting methods applied to IBIS spectra.
In the left panel, the method is applied to a spectrum containing only an absorption component and is therefore fitted with a single Voigt model.
A spectrum exhibiting a complex mix of absorption and emission is fitted with a double Voigt model in the right panel.
Both of these fitted profiles show excellent agreement with the observed profiles, in particular, around the line core where the fitting was prioritised.

\begin{figure}[!t]
	\centering
	\includegraphics[width=0.7\textwidth]{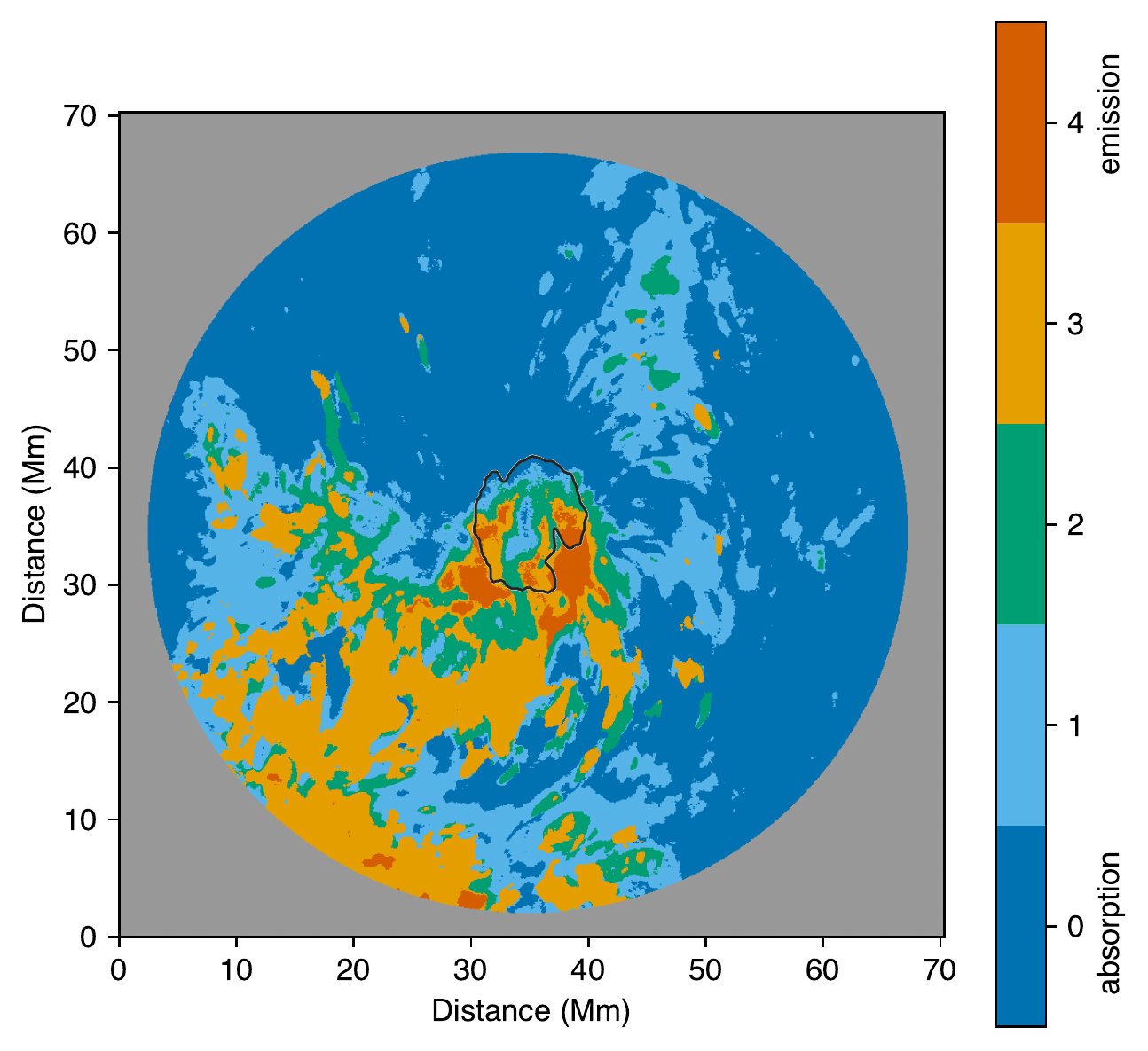}
	\caption{The neural network classifications of the spectra extracted from a single IBIS spectral imaging scan, where the colour bar relates to the spectral shape classified, with `0' and `4' representing pure absorption and emission profiles, respectively. The umbra/penumbra boundary is highlighted using a black contour, and is consistent with that shown in Figure~{\ref{fig:coaligned}}.}
	\label{fig:classificationmap}
\end{figure}

\subsection{Calculating velocities}
Once the quiescent and active components have been isolated, they can be studied independently to ascertain their respective Doppler shifts.
The wavelength of the line core, $\lambda_{\text{observed}}$, for each atmospheric component can be extracted from the parameters defining the central wavelengths of the best-fitting constituent Voigt profiles.
The `at rest' line core wavelength, $\lambda_{\text{stationary}}$, is calculated by applying the algorithm to a spatially and temporally averaged spectrum extracted from a region containing quiet Sun, as documented in Section~\ref{sec:observations} (see, e.g., the rectangular quiet Sun region depicted in Figure~{\ref{fig:coaligned}} and its average profile shown in Figure~{\ref{fig:averagespectrum}}).
Doppler velocities, $v$, can then be calculated by comparing the line core wavelengths of each of the atmospheric components to the stationary wavelength via,
\begin{equation}
	v~(\text{km/s}) = \dfrac{\lambda_{\text{observed}} - \lambda_{\text{stationary}}}{\lambda_{\text{stationary}}} \times 300{\,}000
\end{equation}

\section{Proof of concept testing with IBIS data}
\label{sec:proofofconcept}
\begin{figure}[!t]
	\centering
	\includegraphics[width=\textwidth]{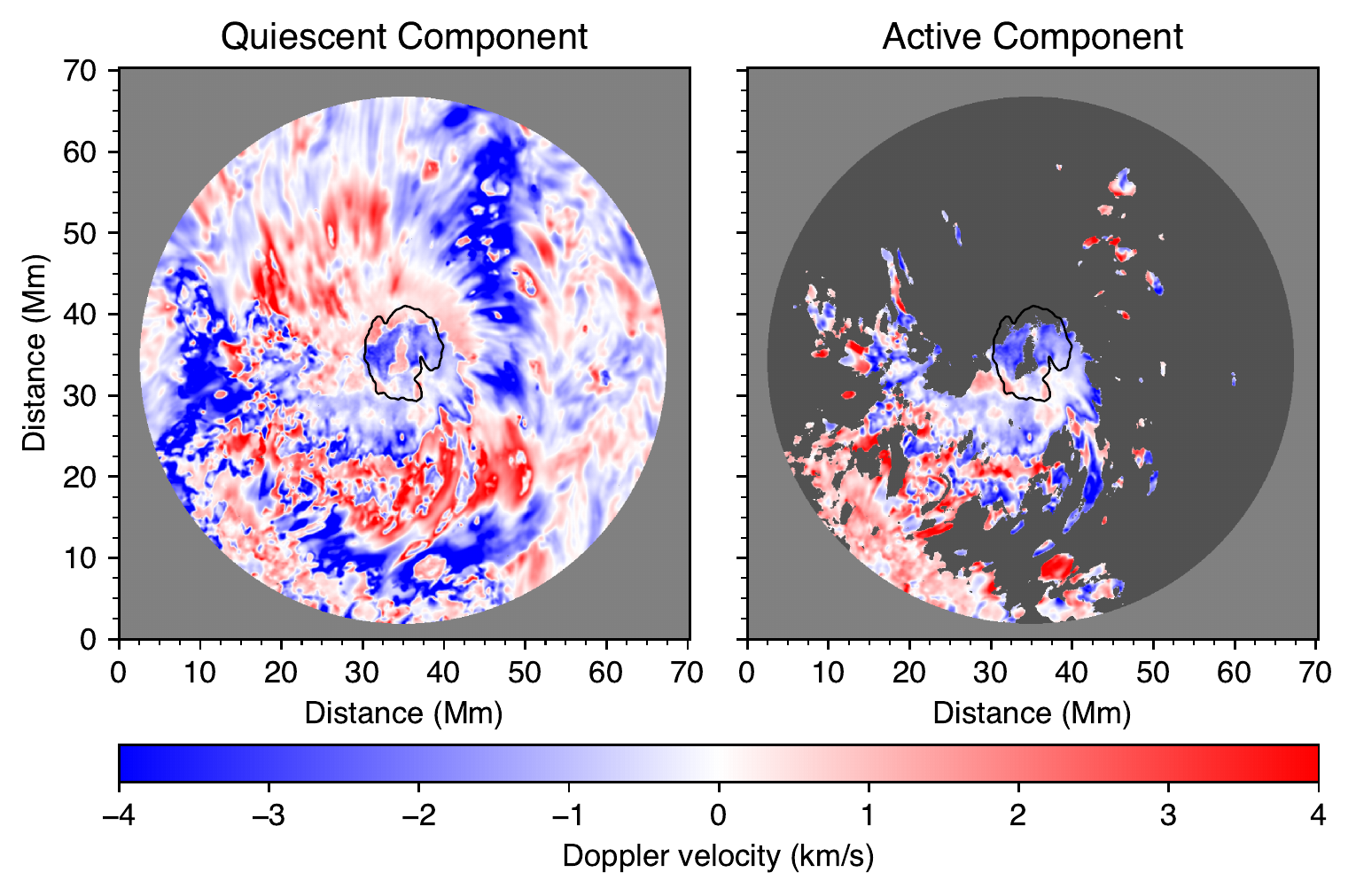}
	\caption{The Doppler velocities corresponding to the fitted absorption components are shown in the left panel, with the colour bar saturated between $\pm$4~km/s to aid visual clarity. For those profiles identified as containing two atmospheric components (i.e., are fitted using a double Voigt model), the Doppler velocities of the secondary emission profile are shown in the right panel. Pixels not requiring a secondary profile fit are shown with a darker grey colour to better isolate them from those pixels requiring a double Voigt model.}
	\label{fig:velocities}
\end{figure}

A proof of concept test was performed using the IBIS dataset described in Section~\ref{sec:observations}.
The spectra across the full field-of-view for a single spectral imaging scan (totalling 664{\,}796 individual spectra) were first classified by the neural network, with the resulting classifications shown in Figure~\ref{fig:classificationmap}.
Following the methods outlined in Section~{\ref{sec:methods}}, Doppler velocities were computed for each fitted absorption component, with the resulting velocity maps shown in the left panel of Figure~{\ref{fig:velocities}}. If a spectral profile required multiple profile fits (i.e., using a double Voigt model), then the resulting Doppler velocities inferred from the emission Voigt fit are shown in the right panel of Figure~{\ref{fig:velocities}}. 
In particular, it can be seen in the right panel of Figure~{\ref{fig:velocities}} that many of the derived Doppler velocities associated with dynamic phenomena appear to rapidly change between neighbouring pixels, an effect that has been documented in previous velocity studies of the solar chromosphere \cite{LohnerBottcher2015, Samanta2016, Kuckein2017}. 
These rapid velocity excursions appear to be closely linked to instances when the {\CaIR} line goes into emission. 
As a result, we believe that such discontinuities may be caused by dynamic changes in the opacity of the plasma, resulting in shifts of the response function of the line caused by the source function no longer monotonically decreasing throughout the chromosphere \cite{delaCruzRodriguez2013, Houston2020}, and hence are not numeric artefacts. 
As such, we must stress that smooth velocity fields should not necessarily be expected following the application of our techniques, especially when examining the challenging effects of radiative transfer in the solar chromosphere.

\begin{figure}[!t]
	\centering
	\includegraphics[width=\textwidth]{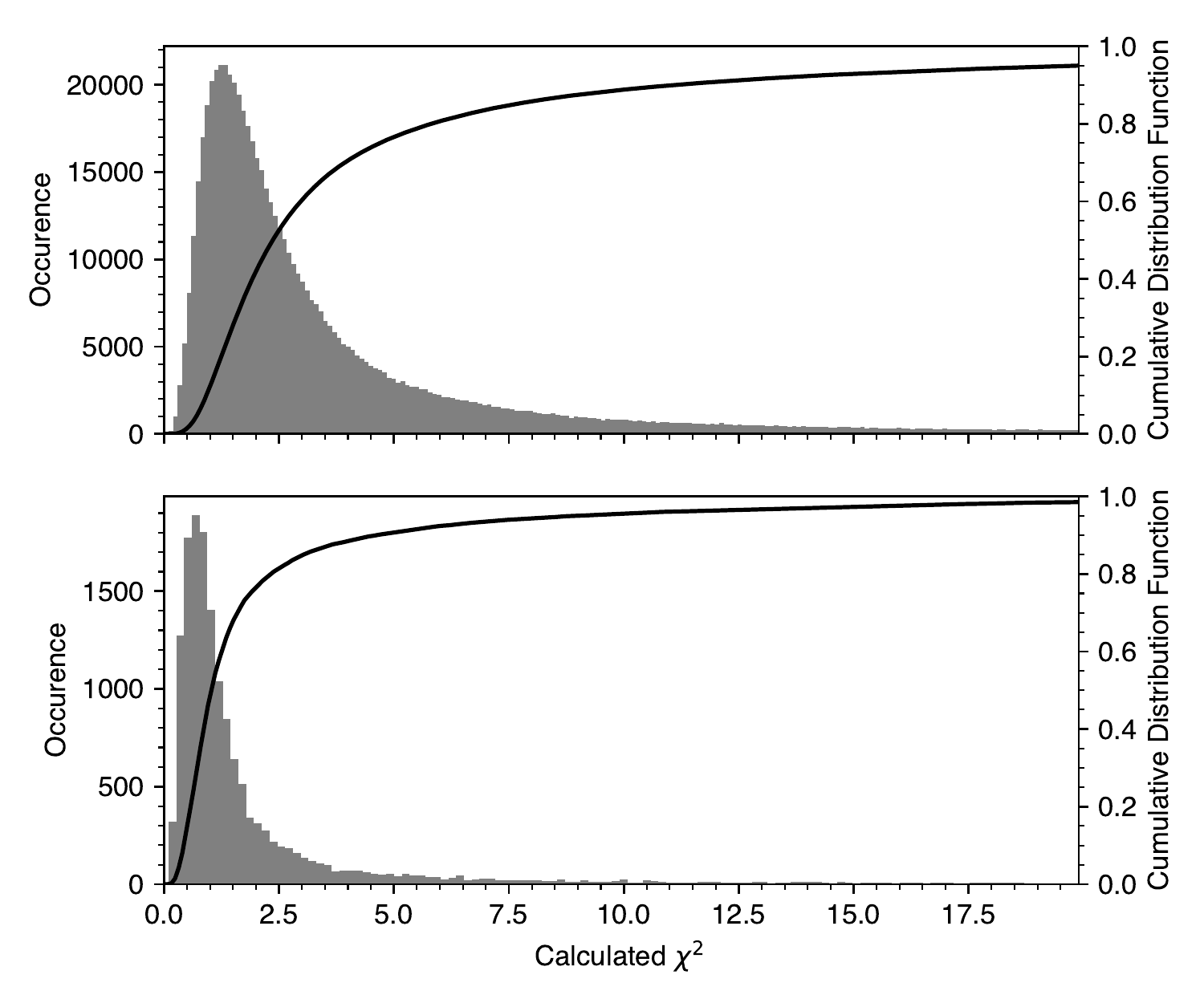}
	\caption{A histogram of occurrences of the modified $\chi^2$ values (derived using Equation~\ref{eq:chi2}) for a single IBIS spectral imaging scan (top panel) is plotted in grey, which depicts the goodness-of-fits between the modelled and observed spectral profiles. In the bottom panel, the histogram is reproduced including only $\chi^2$ values for umbral pixels (as highlighted by the white contours in Figure~{\ref{fig:coaligned}}). In each panel, the solid black lines show the cumulative distribution functions, with the highest 5.0\% and 1.5\% of the calculated $\chi^2$ values omitted from the top and bottom panels, respectively. The median $\chi^2$ values for the entire field-of-view and the umbral locations are $\widetilde{\chi}^{2}=2.36$ and $\widetilde{\chi}^{2}=1.03$, respectively.} 
	\label{fig:chi2}
\end{figure}

A goodness-of-fit value was estimated at each pixel of the field-of-view using a modified $\chi^2$ relation,
\begin{align}\label{eq:chi2}
    \chi^2 = \dfrac{s}{\nu} \sum_{\lambda \in \lambda_c} \dfrac{ \left({I_\lambda^{\text{fitted}}} - {I_\lambda^{\text{observed}}}\right)^2 }{I_\lambda^{\text{observed}}} \ ,
\end{align}
where $s$ is a scaling factor, $\nu$ is the estimated degrees of freedom, and $I_\lambda^{\text{fitted}}$ and $I_\lambda^{\text{observed}}$ are the intensity values of the fitted and observed spectra, respectively.
The wavelengths over which this calculation is performed, $\lambda_c$, include the wavelength closest to the stationary line core, in addition to the 12 wavelength points either side. 
Thus, the central 25 wavelength points (out of 33 points in total) were used to compute the modified $\chi^2$ statistic as this allowed for a better measure of the goodness-of-fit around the line core. 
Therefore, we introduce the scaling factor, $s=\sfrac{\scriptsize{33}}{{\,}\scriptsize{25}}${\,}, to account for this subset of wavelength points. 
The degrees of freedom, $\nu$, were calculated for each spectrum by subtracting the number of fitted parameters (4 for single Voigt profiles and 8 for double Voigt profiles) from the total number of wavelength points, 33.
The $\chi^2$ values for all the fitted spectra in the field-of-view are summarised in Figure~\ref{fig:chi2}.

Considering the entire IBIS spectral imaging scan, the median $\chi^2$ value is $\widetilde{\chi}^{2}=2.36$.
When only the modified $\chi^2$ values for the umbral locations (i.e., spectra within the white umbral contours highlighted in Figure~{\ref{fig:coaligned}}) are used, the median value is $\widetilde{\chi}^{2}=1.03$.
The calculated median $\chi^2$ values are close to one, suggesting that the fitting method is able to accurately constrain the observed spectral line profiles.
The accuracy is particularly good when considering umbral pixels, since not only is the median $\chi^2$ value particularly close to one, as can be seen in Figure~{\ref{fig:chi2}}, but the tail on the $\chi^2$ distribution for umbral spectra drops off very rapidly with increasing $\chi^2$ values.

The most computationally intensive aspect of the proof of concept testing was the fitting of suitable Voigt models to the spectra, with the double Voigt model taking more time than than the single Voigt model.
To fit a full spectral imaging scan totalling 664{\,}796 individual spectra, with 30\% being modelled using a double Voigt profile (see Figure~{\ref{fig:classificationmap}}), took 123~minutes running across all 16-cores on a 2.10~GHz Intel Xeon processor. As a result, processing all 2103 spectral scans from the current IBIS dataset on a single CPU would likely take on the order of 175~days. However, the techniques presented can be further parallelised by employing multiple CPUs, hence bringing the entire processing time down to the order of 1~week or less. Furthermore, as discussed in Section~3b({\ref{sec:mathematics}}), the current algorithms may be ported across to GPUs, providing the ability to accelerate processing performance by an order-of-magnitude or more.

\vspace{-3px}

\section{Discussion}

Although our primary objective for this method is to accurately constrain velocity information within an umbral region, the method can be applied more generally to any region of a spectral imaging dataset where a two-component atmosphere may be present.
As shown in our proof of concept test (Section~{\ref{sec:proofofconcept}}), our methods can be applied to any solar region, including both dynamic locations (umbrae, penumbrae, regions of magnetism, etc.) and those demonstrating more quiescent behaviour (e.g., quiet Sun that is permeated by granulation).  
Although the set of labelled ground truth spectra that are used to train and test the neural network are chosen from within the umbral region, the range of spectral profile shapes encountered within the umbra (i.e., spanning pure absorption through to pure emission characteristics) are representative of many different spectra found outside of the umbral region.

The precision and recall scores of 83\% and 81\%, respectively, as introduced in Section~3c({\ref{sec:IBIStraining}}), suggest that the neural network is able to classify spectra with a reasonable degree of accuracy.
Since our method currently treats classifications `2', `3', and `4' identically, the performance statistics can be recalculated assuming these cases all have the same classification.
This results in increased precision and recall scores of 91\% and 90\%, respectively, suggesting that the performance of the neural network is particularly well-suited for our methods.
Similarly, if we adjust the neural network to distinguish between `emission' (classifications `2', `3', and `4') and `no emission' (classifications `0' and `1'), the precision and recall scores for these groupings are 96\% and 95\%, respectively.
With a larger set of ground truth data, perhaps including other highly dynamical solar phenomena that often exhibit enhanced line-wing asymmetries (e.g., penumbral jets, spicules, magnetic reconnection; \cite{Tian2014, Kuridze2015, Sharma2017}), the precision and recall scores of the neural network could be improved yet further, even to be very close to 100\%.
The precision and recall values computed here are consistent with other trained astrophysical neural networks adopted into mainstream data processing \cite{Dieleman2015,Wright2015}.

The classification methods presented here (i.e., excluding the line fitment and velocity processing) are also useful for estimating the degree of quiescence in a particular dataset \cite{Grant2020}. 
Such processing is carried out by applying only the neural network classification procedure to the data and monitoring the relative occurrence of each of the five classifications --- a process that can be accomplished within a few minutes.

In the future, our methods could be adapted to find velocity measurements for chromospheric jets that often demonstrate plasma motions in the range of $\sim 20-40$~km/s \cite{Pontieu2004}.
Another potential use case is the study of Ellerman bombs, which have been observed in the {\CaIR} line core as significant blue-shifted emission \cite{SocasNavarro2006,Rezaei2015}.
This blue-shifted emission is present in other chromospheric lines, including H$\alpha$ and He~{\sc{i}}~10830{\,}{\AA}, but not in any photospheric lines \cite{Rezaei2015}. As described above, it may be necessary to further train the neural network using a larger set of ground truth data, including examples of enhanced line-wing asymmetries that are synonymous with such dynamical solar phenomena.

Our method could also be extended to model a three (or more) component atmosphere by including additional Voigt (or similar) profiles in the model and modifying the criteria that determine how the assigned classification adjusts the fitting method.
For such a technique to produce accurate velocities, a much higher number of wavelength samples would be required, such that the components are more clearly resolved and are therefore less blended with the surrounding plasma.
With a higher number of wavelength samples, other features, such as the presence of double-peaked self-reversal structures in the emission components, could also be resolved.
To facilitate such future code development, attention will likely need to be turned to the next generation of spectral imaging Fabry-P{\'{e}}rot instruments, in addition to slit- and fibre-based spectropolarimeters. These revolutionary instruments, including the Visible Tunable Filter (VTF; \cite{Schmidt2016}) and the Diffraction Limited Near Infrared Spectropolarimeter (DL-NIRSP), will soon be commissioned the National Science Foundation's Daniel K. Inouye Solar Telescope (DKIST; \cite{Tritschler2016}).

The `active' and `quiescent' components present in Stokes~$I$ observations can be isolated using our method (see, e.g., Figure~{\ref{fig:velocities}}).
Future work to investigate passing these isolated components through the Non-LTE Inversion Code using the Lorien Engine (NICOLE; \cite{SocasNavarro2015}) or the CAlcium Inversion using a Spectral ARchive (CAISAR;\cite{Beck2014}) inversion codes would be of particular interest. 
Often these inversion codes provide plasma parameter outputs based on a static atmosphere. 
Hence, by treating the two atmospheric components separately, ambiguities in the inverted plasma parameters could be minimised, since each component would be better constrained by its own single, high-quality spectral fit. This is a timely endeavour, especially with new, high spectral precision observations on the horizon from new telescope facilities, including DKIST. 

The code is fully available for the community to download and utilise.
Details of how to access it are provided in Appendix~\ref{sec:code}.

\section{Conclusion}

 Novel methods for accurately constraining velocity information from spectral imaging observations have been presented.
 Using machine learning techniques, our methods automatically adapt the spectral models used to fit the input spectra.
 Importantly, our methods will only fit multi-component models if multiple signatures are observed in the input spectra, hence saving time and preventing the overfitment of the data.

 By modelling each atmospheric component with its own independent Voigt function, the constituent components of the atmosphere can be isolated, both spatially and temporally.
 Such techniques have a diverse range of use cases, including the applicability to upcoming DKIST observations, as well as refining the inputs for modern inversion routines, since it enables each atmospheric component to be studied independently.

 A proof of concept test applying this method to a challenging {\CaIR} spectral imaging dataset was presented.
 In this, we demonstrated both the accuracy of the method and how its techniques can be applied more generally.
 Importantly, the algorithms presented are available to the global community through regularly updated download links.

\enlargethispage{20pt}


\dataccess{The data used in this paper are from the observing campaign entitled {\it{`Nanoflare Activity in the Lower Solar Atmosphere'}} (NSO-SP proposal T1020; principal investigator: DBJ), which employed the ground-based Dunn Solar Telescope, USA, during August 2014. The Dunn Solar Telescope at Sacramento Peak/NM was operated by the National Solar Observatory (NSO). NSO is operated by the Association of Universities for Research in Astronomy (AURA), Inc., under cooperative agreement with the National Science Foundation (NSF). Additional supporting observations were obtained from the publicly available NASA's Solar Dynamics Observatory (\href{https://sdo.gsfc.nasa.gov}{https://sdo.gsfc.nasa.gov}) data archive, which can be accessed via \href{http://jsoc.stanford.edu/ajax/lookdata.html}{http://jsoc.stanford.edu/ajax/lookdata.html}. The data that support the plots within this paper and other findings of this study are available from the corresponding author upon reasonable request.}

\aucontribute{CDM and DBJ conceived of and designed the study. DBJ carried out the experiments. CDM performed the data reduction and scientific analysis, with assistance from DBJ, SDTG, PHK, and MS. CDM drafted the manuscript, with theoretical input provided by EK. All authors read and approved the manuscript.}

\competing{The authors declare that they have no competing interests.}

\funding{This work was supported by: 
\vspace{-4mm}
\begin{itemize}
    \item The Department for the Economy (Northern Ireland) through their postgraduate research studentship;
    \item An Invest NI and Randox Laboratories Ltd. Research \& Development Grant (059RDEN-1).
\end{itemize}}

\vspace{-5.5mm}
\ack{CDM would like to thank the Northern Ireland Department for the Economy for the award of a PhD studentship. DBJ and SDTG are grateful to Invest NI and Randox Laboratories Ltd. for the award of a Research \& Development Grant (059RDEN-1) that allowed the computational techniques employed to be developed. The authors wish to acknowledge scientific discussions with the Waves in the Lower Solar Atmosphere (WaLSA; \href{www.WaLSA.team}{www.WaLSA.team}) team, which is supported by the Research Council of Norway (project number 262622), and The Royal Society through the award of funding to host the Theo Murphy Discussion Meeting ``High resolution wave dynamics in the lower solar atmosphere'' (grant Hooke18b/SCTM).}


\appendix{
\section{Availability of the code}
\label{sec:code}
The methods presented in this study have been compiled into a Python package {\it Multi-Component Atmospheric Line Fitting} (MCALF; \cite{MacBride2020})\footnote{\href{https://github.com/ConorMacBride/mcalf}{https://github.com/ConorMacBride/mcalf}}. 
This software package includes a number of subpackages, namely,
\begin{itemize}
    \item \texttt{mcalf.models} --- a collection of models, built upon a common base model, for fitting specific datasets; 
    \item \texttt{mcalf.profiles} --- profile functions that are used to form models;
    \item \texttt{mcalf.visualisation} --- a collection of functions for visualising input and output data; and 
    \item \texttt{mcalf.utils} --- a collection of utility functions used throughout.
\end{itemize}

This package provides a `toolkit' that can be used to define a model optimised for a particular dataset.
A base model is provided that is suitable for any spectral imaging instrument, as well as the custom model derived in the present study that has been optimised for the IBIS sunspot dataset we introduce in Section~\ref{sec:observations}.
The user can easily build upon the base model using our custom template as an example.
This allows the user to apply specific bounds on the fitted parameters and provide their own ground truth dataset of spectral shapes that they would like their neural network to be able to distinguish between.
Additional logic can also be included to interpret their own neural network classifications and take care of the specific spectral shapes present in their dataset.
}




\end{document}